\documentclass[%
 reprint,
 superscriptaddress,
 groupedaddress,
 unsortedaddress,
 runinaddress,
 frontmatterverbose, 
 showpacs,preprintnumbers,
 nofootinbib,
 nobibnotes,
 amsmath,amssymb,
 aps,
 pra,
 prb,
 rmp,
 prstab,
 prstper,
 floatfix,
 ]{revtex4-1}
\usepackage{amsmath}
\usepackage{amssymb}
\usepackage{units}
\usepackage{color}
\usepackage{graphicx}
\usepackage{dcolumn}
\usepackage{bm}
\usepackage{fixmath}
\usepackage{float}
\usepackage{upgreek}
\usepackage{siunitx}

\begin{document}
	
\title{Direct Optimal Control Approach to Laser-Driven Quantum Particle Dynamics}

\author{Alejandro R. Ramos Ramos }
\author{Oliver K\"{u}hn}%
\email{oliver.kuehn@uni-rostock.de}

\affiliation{%
Institute of Physics, University of Rostock, Albert-Einstein-Str. 23-24, 18059 Rostock, Germany
}%
%

\begin{abstract}
Optimal control theory is usually formulated as an indirect method requiring the solution of a two-point boundary value problem. Practically, the solution is obtained by  iterative forward and backward propagation of quantum wavepackets.
Here, we propose direct optimal control as a robust and flexible alternative. It is based on a discretization of the dynamical equations resulting in a nonlinear optimization problem. The method is illustrated for the case of laser-driven wavepacket dynamics in a bistable potential. The wavepacket is parameterized in terms of a single Gaussian function and field optimization is performed for a wide range of particle masses and lengths of the control interval. Using the optimized field in a full quantum propagation still yields reasonable control yields for most of the considered cases. Analysis of the deviations leads to conditions which have to be fulfilled to make the semiclassical single Gaussian approximation meaningful for field optimization.
\end{abstract} 

\maketitle

\section{Introduction}

``Teaching lasers to control molecules'' has been a long-standing goal in molecular physics~\cite{judson92_1500}.
Among the various methods of the early days~\cite{paramonov83_340,tannor85_5013,tannor86_5805,brumer86_541,judson92_1500}, optical control theory  OCT) emerged as a versatile tool. Originally developed by Rabitz et al. \cite{shi88_6870,shi89_185} and Kosloff et al. \cite{kosloff89_201}, numerous methodological extensions have been developed over the years (for reviews, see e.g. \cite{brif10_075008,werschnik07_R175,worth13_113,keefer18_2279}). In terms of practical realizations of chemical reaction control, the feedback strategy~\cite{judson92_1500,brixner03_418,prokhorenko06_1257} as well as straightforward resonant excitation schemes~\cite{stensitzki18_126,nunes20_8034,heyne19_11730} have been most successful.

In quantum optimal control theory the goal of optimizing the expectation value of a target operator such as a projector onto a certain state,  is formulated as a variational problem for a cost functional subject to certain constraints. The latter include, for instance, some penalty for high field intensities or that the wavepacket should fulfill the Schr\"odinger equation. This control problem is usually solved using an \textit{indirect} approach, i.e.\ the cost functional is not minimized directly. Instead, the stationarity condition for the cost functional is converted to a two-point boundary problem for two coupled Schr\"odinger equations. A numerical solution is obtained by iterative forward and backward propagation of the actual wavepacket and an auxiliary wavepacket, respectively (e.g. \cite{zhu98_385}).
This procedure is sometimes referred as the optimize and then discretize paradigm \cite{kelly17_849}. Indirect methods for optimal control are in use in other areas of physics, e.g. stochastic control~\cite{kappen07_149}, but also in engineering and biology~\cite{chen-charpentier20_112983}.

\textit{Direct} optimal control, in contrast, follows the  discretize and then optimize paradigm, i.e.\  the cost functional is minimized directly using methods from nonlinear optimization. Although being popular, for instance, in applied mathematics~\cite{betts10_}, engineering~\cite{pardo16_946}, and biology~\cite{chen-charpentier20_112983}, there have been no applications to quantum molecular dynamics so far. The present paper is  devoted to fill this gap.

Indirect optimal control requires to solve iteratively two time-dependent Schr\"odinger equations where the numerical effort scales exponentially with the number of degrees of freedom. To cope with this situation the Multi-Configurational Time-Dependent Hartree (MCTDH) approach is most suited~\cite{meyer90_73,beck00_1}. An OCT implementation has been reported in Ref.~\cite{schroder08_850}, for an application see also Ref.~\cite{accardi09_7491}. The solution of the time-dependent Schr\"odinger equation requires a priori knowledge of the potential energy surface. But, when driving the wavepacket into a particular region of configuration space using laser control, a global potential might not be needed.
 Thus on-the-fly approaches, e.g, in the context of MCTDH~\cite{richings15_269,richings18_134116} could be of advantage. On the other hand, semiclassical approximations in terms of Gaussian wavepackets play a prominent role in molecular quantum dynamics~\cite{heller18_} and indeed there has been a semiclassical formulation of indirect OCT reported in Refs.~\cite{kondorskiy05_75,kondorskiy05_041401} (for related work using Wigner space sampling, see Ref. \cite{bonacic-koutecky05_11}).

In this paper we explore direct OCT using a representation of the wavepacket dynamics in terms of a single Gaussian function. Although this choice has been made for numerical convenience, it also facilitates exploration of its limitations by comparison with solutions of the time-dependent Schr\"odinger equation. Specifically, for the considered problem of quantum particle motion in a bistable potential we are able to identify conditions for which the single Gaussian approximation is adequate.

\section{Theoretical Methods}
\label{sec:Theory}
\subsection{Equations of motions}
\label{sec:eom}
%
The equations for the time evolution of a quantum mechanical state can be obtained from the TDVP starting with the stationarity condition for the action $S$, i.e.~\cite{broeckhove88_547}
\begin{equation}
  \label{eq:dS=0}
  \delta S = \delta \int_{t_1}^{t_2} L(\Psi,\Psi^*)dt=0~,
\end{equation}
where the quantum Lagrangian is given by (Note that atomic units are used throughout)

\begin{equation}
  L=\left\langle \Psi \left| i \frac{\partial}{\partial t} -{H}(t)   \right| \Psi \right \rangle \,.
\end{equation}
In the following we will focus on one-dimensional systems (coordinate $x$ and momentum $p$) coupled to a radiation field, $E(t)$, in dipole approximation (dipole operator ${\mu}(x)$). Thus the Hamiltonian operator in the coordinate representation is given by
\begin{equation}
  H(t)=H_0 +H_{\rm f}(t) = -\frac{1}{2 m} \frac{d^2}{dx ^2} + {V}(x) -  {\mu}(x) E(t) \,.
\end{equation}
Equation \ref{eq:dS=0} yields the condition~\cite{broeckhove88_547}
\begin{equation}
  \label{tdvp}
  {\rm Re} \left[ \left \langle \delta \Psi \left| i  \frac{\partial}{\partial t} - H(t)   \right| \Psi \right \rangle  \right] =0~.
\end{equation}
Assuming  that the wavepacket is parameterized by the set of parameters ${\bm a}=\{a_1,...,a_n\}$ this yields
\begin{equation}\label{wf_variation}
  \delta \Psi = \sum_{j=1}^n \left( \frac{\partial \Psi }{\partial a_j}\right) \delta a_j \, .
\end{equation}
Inserting equation (\ref{wf_variation}) into equation (\ref{tdvp}) gives the equations of motion for the general set of parameters used to describe the wavepacket
\begin{equation}\label{dot_alpha}
  \dot{a_i}=-  \sum_{j=1}^n K_{ij} \,{\rm Re}\left \langle \frac{\partial \Psi }{\partial a_j}  \Big| {H} \Psi \right\rangle \quad \forall i=1,\ldots,n~,
\end{equation}
with $K_{ij}$ being the elements of the inverse of the matrix formed by ${\rm Im}\langle {\partial \Psi }/{\partial a_i} | {\partial \Psi }/{\partial a_j} \rangle$.

For the purpose of illustration we  assume that the wavepacket has the following Gaussian form~\cite{heller18_} %
\begin{equation}
\label{eq:psi_gauss}
  \Psi(x,\alpha,\beta,x_0,p_0)=\left(\frac{2 \alpha}{\pi}\right)^{1/4} \exp\left[-(\alpha+i\beta)(x-x_0)^2 + i p_0(x-x_0)\right]~,
\end{equation}
where $\alpha$ and $\beta$ are real numbers describing the width and tilt of the phase space  Gaussian. Further, $x_0$ and $p_0$ are the average position and momentum, respectively. At this point it is important to notice  that the time-dependence of the wavepacket is implicit in the time-dependence of the parameters $\alpha(t)$, $\beta(t)$, $x_0(t)$ and $p_0(t)$. Hence, we identify ${\bm a}={\bm a}(t)=\{\alpha,\beta,x_0,p_0\}$ and using equation (\ref{dot_alpha}) gives the following set of coupled differential equations
\begin{equation}\label{eq:alpha}
\dot{\alpha}=\frac{4\alpha \beta}{m}~,
\end{equation}
\begin{equation}\label{eq:beta}
\dot{\beta}= -\frac{2 (\alpha^2-\beta^2)}{m} - 4\alpha^2 \frac{\partial}{\partial \alpha} U(t)~,
\end{equation}
\begin{equation}\label{eq:x0}
\dot{x}_0= \frac{p_0}{m}~,
\end{equation}
\begin{equation}\label{eq:p0}
\dot{p}_0=- \frac{\partial}{\partial x_0} U(t) \,
\end{equation}
subject to some initial conditions at time $t_{\rm 0}$.
Here, we defined the time-dependent expectation value of the  potential
\begin{equation}
  \label{eq:mfpot}
  U(t) =\langle \Psi (t) | {V}(x) -  \mu(x) E(t)| \Psi(t) \rangle~.
\end{equation}
In the next section we will focus on the control problem assuming that these equations of motion can be solved, which implies that the expectation value of the  potential and its derivatives are available.

\vspace*{0.5cm}
%
\subsection{Statement of the Control Problem}

Let us start with a brief summary of optimal control theory \cite{werschnik07_R175,worth10_15570,brif10_075008}.
Given a functional of the form
\begin{equation}\label{eq:J}
  {\mathcal J}[\bm a, \bm u, \bm k]= {\mathcal T}[\bm a(t_{\rm f}),\bm k,t_{\rm f}] + \int_{t_0}^{t_{\rm f}} {\mathcal R}[\bm a(t),\bm u(t),\bm k,t]\,dt~.
\end{equation}
where ${\mathcal T}$ and ${\mathcal R}$ are the terminal and running cost, respectively, the task is to find the state trajectory $\bm a(t)$, external control $\bm u(t)$ (where the time  $t\in [t_0,t_{\rm f}]$) and the set of static parameters $\bm k$ that minimize the functional ${\mathcal J}[\bm a, \bm u, \bm k]$.
The minimization is performed subject to the following differential constraints
\begin{equation}\label{eq:diffcon}
  \dot{\bm a}(t)=f[\bm a(t),\bm u(t),\bm k,t]~, ~~~t\in [t_0,t_{\rm f}] \, .
\end{equation}
Further, there can be path constraints
\begin{equation}
  \bm h_{\rm L} \leq \bm h[\bm a(t),\bm u(t),\bm k,t]\leq \bm h_{\rm U}~,
\end{equation}
and event constraints such as
\begin{equation}
\bm e_{\rm L} \leq \bm e[\bm F[\bm a(t),\bm u(t)],\bm k,t_0,t_{\rm f}]\leq \bm e_{\rm U}~.
\end{equation}
Here, the subscript ${\rm L}$ and and ${\rm U}$ denotes the lower and upper boundary, respectively, defining the constraints. Notice that in contrast to path constraints, event constraints are not time-dependent, but could include a functional, $\bm F$, of, e.g., the state trajectory or the external control (see below).

Next, we specify this general control problem to the model introduced in section \ref{sec:eom}.
The state is characterized by the set ${\bm a}(t)=\{\alpha,\beta,x_0,p_0\}$ and the external control is given by the laser field $\bm u(t)=E(t)$. Additional time-independent parameters, ${\bm k}$, will not be used. The differential constraints  (\ref{eq:diffcon})   are given by the equations (\ref{eq:alpha}-\ref{eq:p0}).

The goal of the optimization can be stated as follows. Given some initial quantum state $|\Psi(t_{\rm 0})\rangle$, parameterized by $\bm a^{\rm i}=\{\alpha^{\rm i},\beta^{\rm i},x_0^{\rm i},p_0^{\rm i}\}$, find a laser field $E(t)$ such that the overlap is maximized between the time-evolved final state at $t=t_{\rm f}$, $|\Psi(t_{\rm f})\rangle$, and some target state $|\Phi^{\rm t}\rangle$. Thus, the terminal cost in equation (\ref{eq:J}) is given by (notice the minus sign because the terminal cost will be minimized and we want to maximize the overlap)
\begin{equation}
	{\mathcal T}[\bm a(t_{\rm f}),t_{\rm f}] = -\left| \langle  \Psi(t_{\rm f}) | \Phi^{\rm t} \rangle \right|^2
\end{equation}
Here, for simplicity we will use the parametrization of equation (\ref{eq:psi_gauss}) for the target state as well, labeling the target parameters as $\bm a^{\rm t}=\{\alpha^{\rm t},\beta^{\rm t},x_0^{\rm t},p_0^{\rm t}\}$.

The running cost will be chosen as follows
\begin{equation}\label{run_cost}
{\mathcal R}[E(t),t] =\kappa \frac{ |E(t)|^2}{s(t)}~,~~~\quad s(t)=\sin^2 \left( \frac{\pi}{t_{\rm f}} t\right) +\epsilon \, .
\end{equation}
Besides the field intensity we have included a factor $\kappa$ scaling the penalty for high field strengths as well as a shape function $s(t)$, which  ensures that the field increases(decreases) slowly when turned on(off) \cite{sundermann00_1896}. Note that $\epsilon$ is a small parameter introduced to avoid division by zero and numerical problems at times $t=0$ and $t=t_{\rm f}$. Throughout the text we have used $\epsilon=0.005$.

For the application presented below we don't use any path constraints, but  event constraints. Given the event
\begin{equation}\label{event}
\bm e[F[E(t)],\bm a(t_0)]=
  \begin{pmatrix}
    \alpha(t_0) \\
    \beta(t_0) \\
    x_0(t_0) \\
    p_0(t_0) \\
    \int_{t_0}^{t_{\rm f}} E(t) dt
  \end{pmatrix}~,
\end{equation}
upper and lower bounds will be chosen equal as follows
\begin{equation}
  \bm e_{\rm L}= \bm e_{\rm U}=
  \begin{pmatrix}
    \alpha^{\rm i} \\
    \beta^{\rm i} \\
    x_0^{\rm i} \\
    p_0^{\rm i} \\
    0\\
  \end{pmatrix}~.
\end{equation}
Hence, the parameters of the initial state are fixed and not subject to optimization. Further, we enforce the zero-net-force condition by demanding that $F[E(t)]= \int_{t_0}^{t_{\rm f}} E(t) dt=0$~\cite{doslic06_013402}.

The optimization problem will be solved using a direct method, i.e. by means of discretization of the differential equations. Details will be specified in the next section.

\vspace*{0.5cm}

\subsection{Model System and Computational Details} 
\label{sec:model}
The direct optimal control approach will be applied to the problem of particle dynamics in a bistable potential. This could represent, for instance, proton or hydrogen atom transfer in a tautomerization reaction~\cite{doslic98_292, doslic11_411}. The following potential will be used
\begin{equation} \label{eq:pes}
  V(x)=V_{\rm B}\left(\left(\frac{x}{x_{\rm B}}\right)^2-1\right)^2\,.
\end{equation}
Here,  $x_{\rm B}$ is the distance between the minimum of the potential and the top of the barrier, and $V_{\rm B}$ is the barrier height.

\begin{figure}[htb!]
  \centering
  \includegraphics[width=\linewidth]{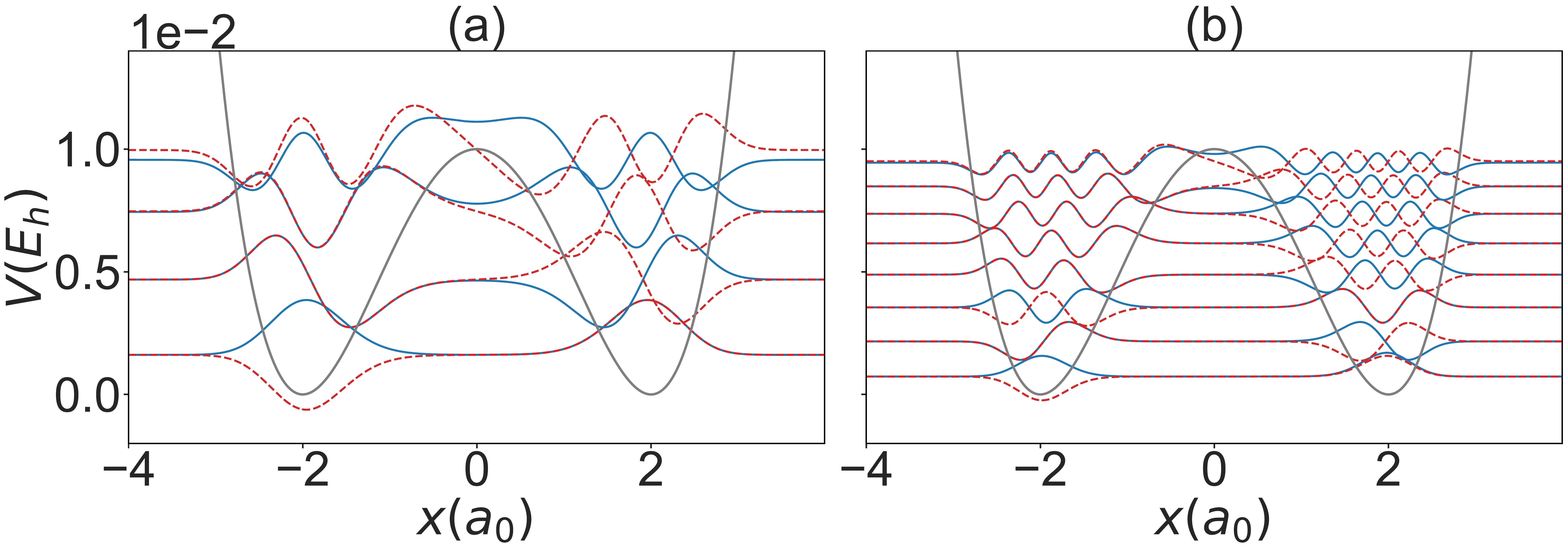}
  \caption{Eigenstates for a particle of mass (a) $1~m_{\rm H}$ and (b) $5~m_{\rm H}$ in the potential given by equation (\ref{eq:pes}) with $x_{\rm B}=2 a_{0}$ and  $V_{\rm B}=0.01 E_{\rm h}$. Solid and dashed lines correspond to even and odd eigenstates, respectively.}\label{fig:eigenstates}
  \end{figure}

The system-field interaction is treated in semiclassical approximation, taking the polarization of the field in the same direction as the dipole, and assuming a linear model for the latter ($q$ is the charge)
\begin{equation}\label{eq:dipole}
\mu(x)=q x\, .
\end{equation}
Specific parameters for the numerical simulations have been chosen to mimic typical situations in proton transfer reactions~\cite{doslic98_292, doslic11_411}, i.e.  $x_{\rm B}=2 a_{0}$ ($\approx 1.06 ~\si{\angstrom})$, $V_{\rm B}=0.01 E_{\rm h}$ $(\approx 6.3$ kcal/mol), and $q=1$ ($=1 e$). The particle's mass, $m$, will be used to tune the `quantumness' of the dynamics. Exemplary, we show potential and eigenstates for two choices of the masses in figure \ref{fig:eigenstates}. Comparing the two cases we note that in particular the number of eigenstates below the barrier is 8 and 16 for masses of $1 ~m_{\rm H}$  and $5~m_{\rm H}$ respectively (where $m_{\rm H}$ is the hydrogen mass).

Using equations (\ref{eq:pes}) and (\ref{eq:dipole}) together with equation (\ref{eq:psi_gauss}) one can calculate the time-dependent expectation value of the  potential, equation (\ref{eq:mfpot}), and its derivatives with respect to $\alpha$ and $x_0$ required for the equations of motion (\ref{eq:beta}) and (\ref{eq:p0}). To this end the potential is  globally approximated by a sum of Gaussians of the form
\begin{equation}\label{eq:pot_approx}
  V(x)\approx \sum_{p=1}^g g_p e^{-b_p(x-x_p)^2}~.
\end{equation}
We have used $g=5$ which gives
$ g_p=\{31.000,-1.529,-1.529,31.000,1.348\}$ (in units of $V_{\rm B}$),
$b_p = \{1.397,1.658,1.658,1.397,0.\}$ (in units of $x_{\rm B}^{-2}$),
and
$
x_p = \{-2.981,-1.142,1.142,2.981,0.\}$ (in units of $x_{\rm B}$).

Using equation (\ref{eq:pot_approx}) one obtains
\begin{equation}\label{V}
 U(t) = \sum_{p=1}^5 g_p e^{-B_p} \left( \frac{2\alpha(t)}{2\alpha(t) +b_p}\right)^{1/2} -q x_0(t) E(t)~,
\end{equation}
\begin{equation}\label{dV_da}
  \frac{\partial }{\partial \alpha} U(t)= \sum_{p=1}^5 D_p \left( \frac{1}{4 \alpha(t)^2} - \frac{b_p}{\alpha(t)(2\alpha(t) +b_p)} (x_0(t)-x_p)^2  \right)~,
\end{equation}
\begin{equation}\label{dV_dx}
  \frac{\partial }{\partial x_0} U(t) = -2\sum_{p=1}^5 D_p (x_0(t)-x_p) - q E(t)~,
\end{equation}
where
\begin{equation}\label{bp_dp}
  B_p=\frac{2\alpha(t) b_p}{2\alpha(t) + b_p}(x_0(t) - x_p)^2
\end{equation}
and
\begin{equation}
	D_p=g_p b_p e^{-B_p} \left( \frac{2\alpha(t)}{2\alpha(t) + b_p}  \right)^{3/2} \,.
\end{equation}

For the solution of the control problem the software package PSOPT has been used~\cite{becerra10_1391}. It provides different discretization schemes. The global pseudospectral Legendre and Chebyshev discretization yield very slow convergence for non-smooth functions \cite{kelly17_849}, as it is the case for the solutions found for $\alpha(t)$ and $\beta(t)$ (see first and second row, (b) and (d) columns of figure \ref{fig:guess_solution} below). Increasing the number of nodes is not an option for these discretization schemes because of the non-sparsity of the Jacobian matrices which cannot be handled properly by the implemented IPOPT NLP (nonlinear programming) solver \cite{wachter06_25}. This issue translates into a disproportional increase of computational time.
The local  methods available are trapezoidal and Hermite-Simpson discretization. In order to check their performance we simulated the case of  a particle of mass of $1~m_{\rm H}$ and a final time of $t_{\rm f}=20000~{\rm au}$. In doing so the number of time discretization nodes  has been scanned  from 200 to 6000.  To evaluate the discretization error we use the maximum relative local error, $\varepsilon_{\rm disc}$,  defined in reference \cite{becerra10_1391}.  The results are shown in figure \ref{fig:collocation}. If the number of nodes is below 1000 the trapezoidal method has a smaller error $\varepsilon_{\rm disc}$ compared to Hermite-Simpson for the same number of nodes. Beyond 1000 nodes, Hermite-Simpson outperforms the trapezoidal discretization. However, this comes at the expense of an increased computational time as can be seen in the lower panel of figure \ref{fig:collocation}.  For the simulations reported below we have used Hermite-Simpson discretization with 2000 nodes, which offers a good balance between accuracy and speed.
\newcommand{\figwidth}{0.7}
\begin{figure}[hbtp]
\centering
        \centering
      \includegraphics[width=\figwidth\linewidth]{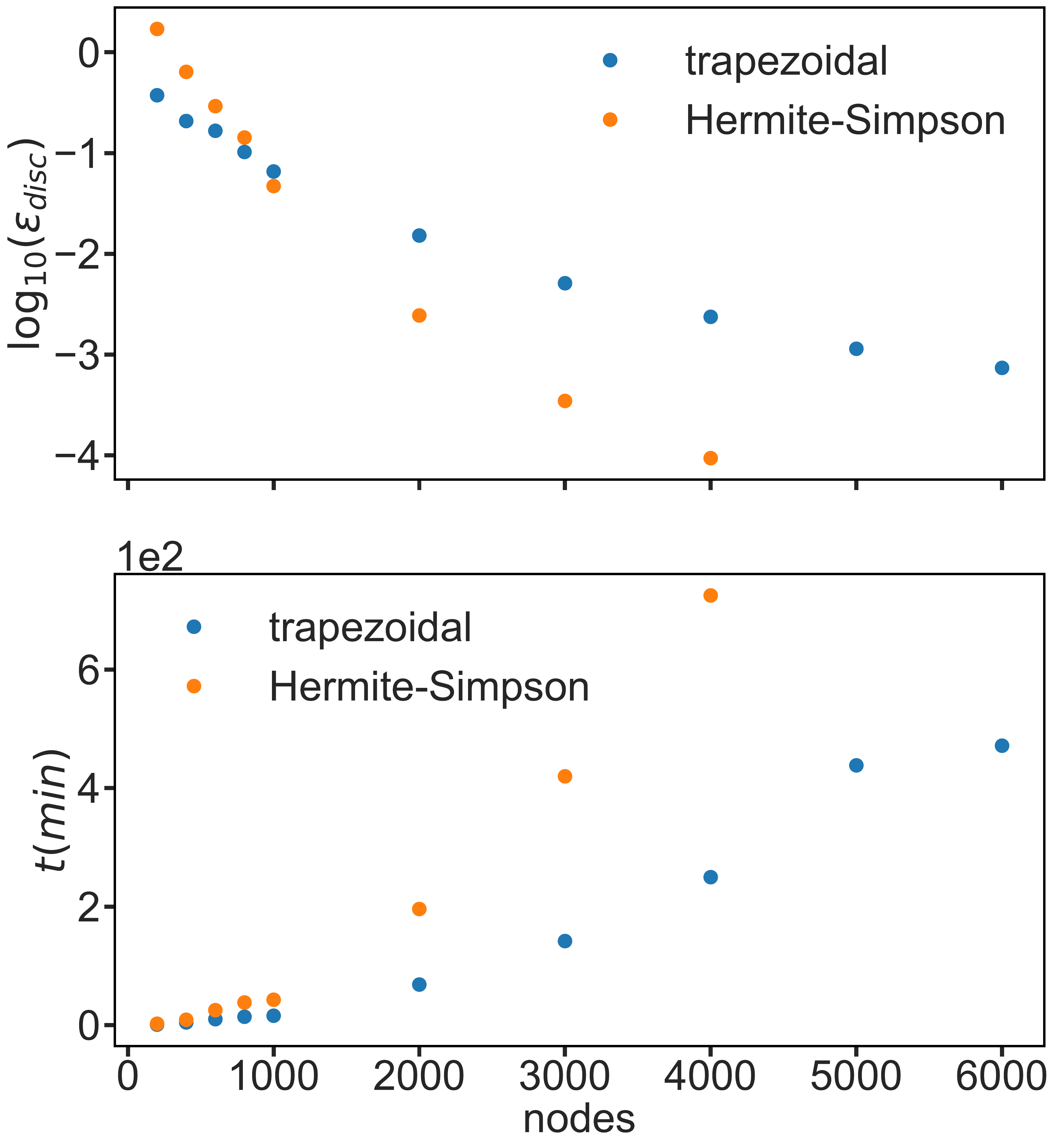}
        \caption{Maximum relative local error (upper panel) and timing (lower panel) for trapezoidal (blue) and Hermite-Simpson (orange) as a function of the number of nodes. }\label{fig:collocation}
\end{figure}

In order to quantify the importance of quantum effects beyond the simple Gaussian ansatz for the wavepacket, equation (\ref{eq:psi_gauss}), MCTDH simulations have been performed using the optimized field. For this purpose the Heidelberg MCTDH package has been used~\cite{mctdh85}.
\section{Results}
\subsection{Laser-controlled Proton Transfer}\label{sec:control_over}
In the following we present a proof-of-principle application of direct OCT using the example of proton transfer in a bistable potential. Specifically, the two cases (particle masses) given in figure \ref{fig:eigenstates} will be  considered.
For the  initial state we choose the parameters of a Gaussian in the left well, and as the target state we choose a symmetrically located Gaussian in the right side well. The Gaussian parameters have been optimized to the ground state using a local harmonic approximation. Although direct control in principle allows to vary the final time, in the present application the  final time has been fixed to $t_{\rm f}=20000~{\rm au}$. The penalty factor has been chosen as  $\kappa=0.3$ (cf. equation \ref{run_cost}).
To solve the problem we also have to provide an initial guess for states and control which is shown in figure \ref{fig:guess_solution}(a,c). The rapid oscillations have been chosen randomly; there is no correlation between the different variables.

The optimal solutions for the two particle masses are given in figure \ref{fig:guess_solution}(b,d). Apparently, the optimal field is able to drive the center of the wavepacket across the barrier into the right minimum at $t=t_{\rm f}$. In this respect one should note that the optimal fields have a relatively simple shape and little resemblance with the initial guess. This is one of the major advantages of the direct approach to optimal control problems, i.e. the convergence region of the initial guess is very broad. The dynamics is rather similar, i.e. in both cases the trajectory passes the barrier coming from the turning point at the left hand side. Just before and after the barrier the wavepacket gets localized in coordinate and delocalized in momentum space, whereas the position-momentum correlation ($\beta$) vanishes. The wavepacket passes the top of the barrier with large momentum.

\renewcommand{\figwidth}{1}

\begin{figure}[thbp]
\centering
        \centering
        \includegraphics[width=\figwidth\linewidth]{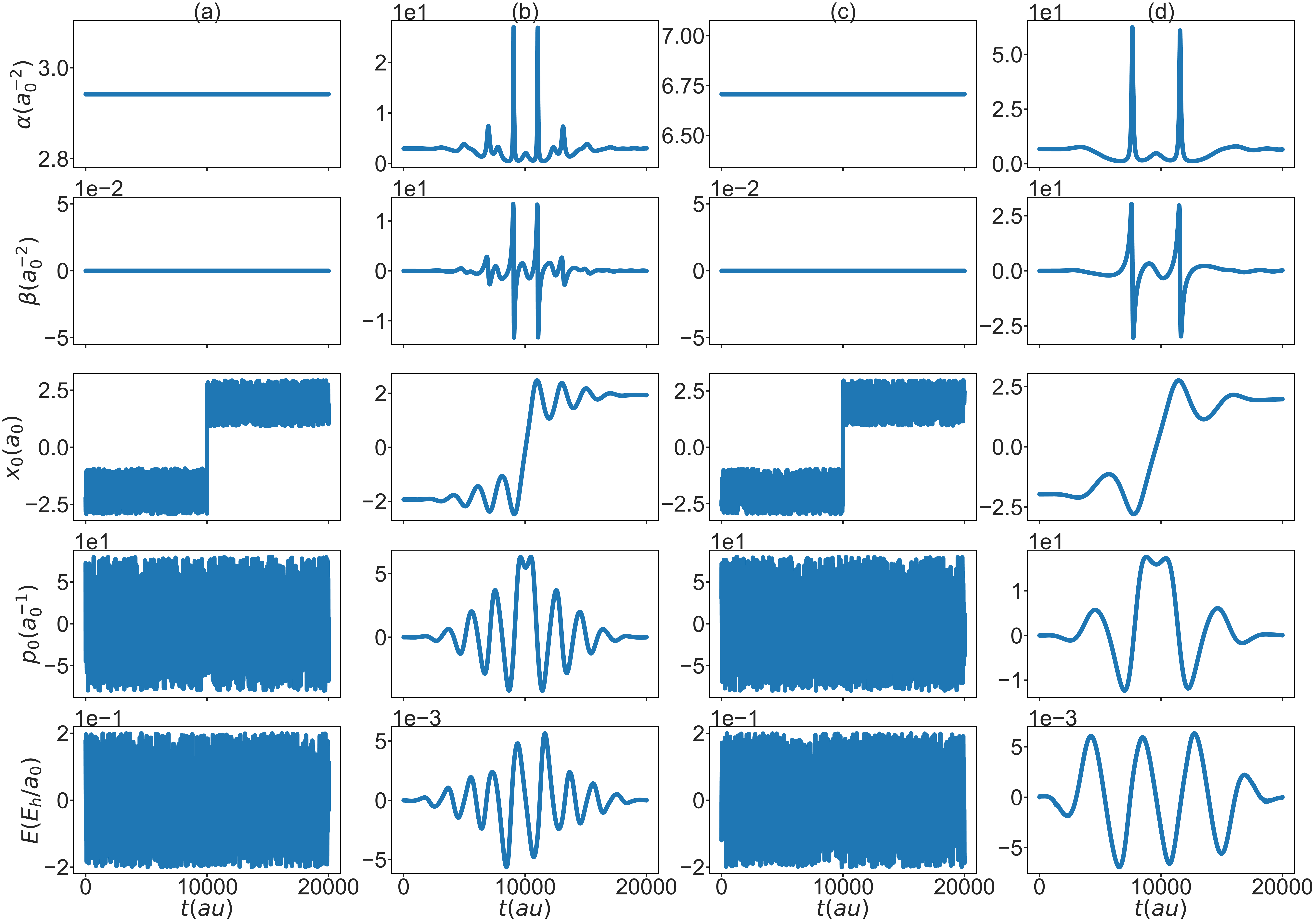}
        \caption{Initial guess (a,c) and optimal solution (b,d)  for state, ${\bm a}$, and control field for two different particle masses ($1~m_{\rm H}$ -- (a,b), $5~m_{\rm H}$ -- (c,d)).}\label{fig:guess_solution}
\end{figure}

\begin{figure}[hbtp]
\centering
        \centering
        \includegraphics[width=0.9\linewidth]{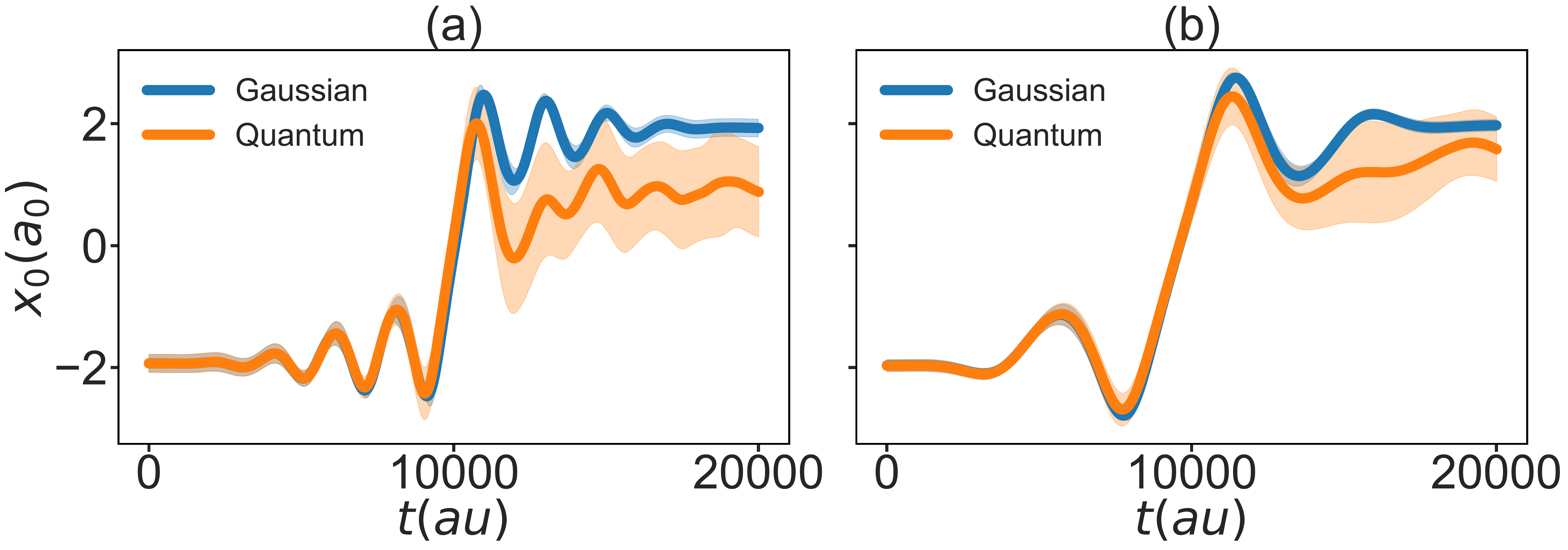}
        \caption{Comparison of the coordinate expectation value and its variance using the Gaussian approximation (blue) and the full quantum propagation (orange), both under the influence of the optimal control field as obtained for the Gaussian ((a) $1~m_{\rm H}$, (b) $5~m_{\rm H}$) .}\label{fig:psopt_mctdh_comp}
\end{figure}

The question now arises if the optimum field found for a single Gaussian wavepacket  is able to trigger the same particle dynamics  in the  full quantum case. To this end the optimal field is used within a quantum dynamics simulation. The results are compared in
figure \ref{fig:psopt_mctdh_comp} in terms of coordinate expectation values and variances. Until after the barrier crossing, Gaussian and full quantum results are rather similar. Indeed, if the goal would have been to trigger the localization of the wavepacket somewhere in the   region of the right well at a particular time, the optimal field would still perform this task also in the quantum case. Of course, the agreement between classical and quantum propagation is better in case of the heavier mass even though there is considerable larger spread of the wavepacket in the quantum case after reflection at the right turning point. For the lighter mass the agreement after barrier crossing is less favorable due to the larger spread and the structured character of the quantum wavepacket which cannot be captured by a single Gaussian.

\vspace*{0.5cm}
\subsection{Region of Validity of the Gaussian Wavepacket Approximation}
Single Gaussians cannot capture the dynamics of structured wavepackets. Nevertheless, the agreement between Gaussian and full quantum results is at least qualitative, even for the lighter particle. This provides the motivation for the investigation of the validity of the Gaussian approximation over a wider range of parameters. Again the optimum field is obtained following the procedure described in section \ref{sec:control_over}, but now for different  final times (ranging from $5000~{\rm au}$ to $20000~{\rm au}$ in steps of $1000~{\rm au}$) and masses (ranging from 1 $m_{\rm H}$ to 10 $m_{\rm H}$ in steps of 1 $m_{\rm H}$). To evaluate the performance  of the optimum field to drive the wavepacket to the right well in the full quantum case we choose the following error:
\begin{equation}\label{eq:error}
  {\rm Err}=\frac{\left|x_0^{\rm t}-\langle {\tilde \Psi}(t_{\rm f}) | x | {\tilde \Psi}(t_{\rm f}) \rangle \right|}{x_{\rm B}}~,
\end{equation}
where $\tilde \Psi(t_f)$ is the exact quantum wavefunction at the final time. This error will be between 0 and 1 if the expectation value of the quantum wavepacket crossed the barrier and greater than 1 if it did not. Results are shown in figure \ref{fig:error}.

\renewcommand{\figwidth}{0.6}
\begin{figure}[th]
\centering
        \includegraphics[width=\figwidth\linewidth]{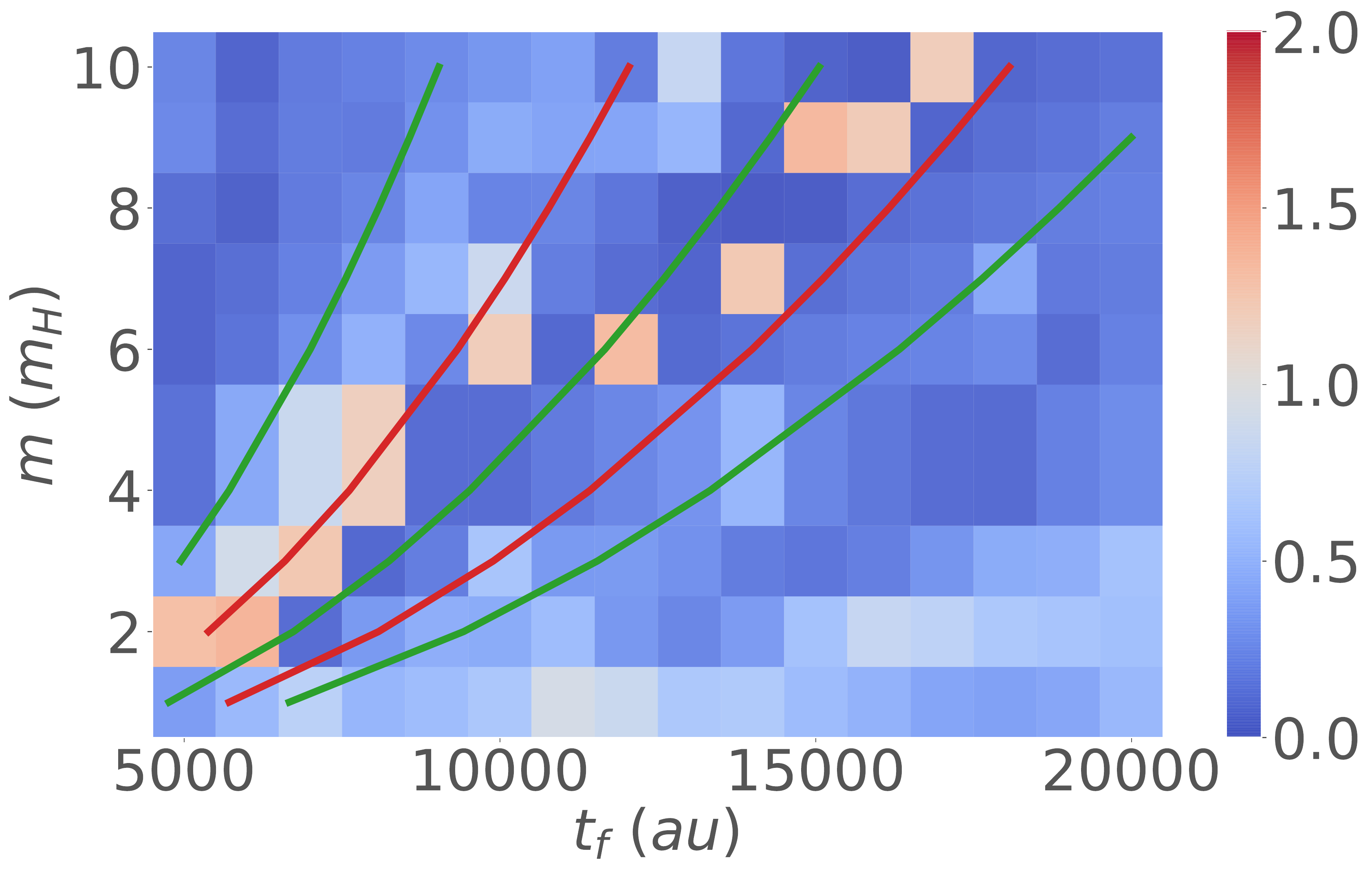}
        \caption{Error according to equation \ref{eq:error} as a function of different final times and masses. Green lines represent an odd number of half harmonic oscillation periods for the corresponding mass $(2n+1) T/2$ with $n=1,2,3$ and red lines represent an integer number of  periods $n T$ with $n=2,3$.}\label{fig:error}
\end{figure}

In general, we can see from figure \ref{fig:error} that the Gaussian optimal control fields  are able to drive the particle reaction on a broad range of masses and final times. As expected the performance deteriorates  for  the lighter masses.
There are some features which deserve closer attention. For example, there are regions where the Gaussian wavepacket approach works exceptionally well (characterized by stripes of intense blue color). In these regions the final time is matching a total integer number of  well oscillations plus the barrier crossing time. Assuming that these oscillations are harmonic with period $T$ and taking the barrier crossing time as being half of the harmonic period, these final times can be estimated. The middle green line in figure \ref{fig:error} corresponds to a final time of $5T/2$. It nicely matches  with the dark blue region where the approach works well. Thus, in general one would expect  regions with $(2n+1) T/2$ and $nT$ where the approximation works well and not so well, respectively. This is roughly seen in figure \ref{fig:error}, although the deviation from the harmonic approximation causes some quantitative disagreement. This analysis points to the importance of the final time $t_{\rm f}$ for the effect of the quantumness of the dynamics on the overlap with the target. In passing we note that in principle direct optimal control offers the possibility to optimize the final time as well, e.g., to fulfill some constraints with respect to the spread of the wavepacket.

Another interesting feature apparent from figure \ref{fig:error} are the isolated ``islands'' of poor performance, e.g. at $t_f=14000~{\rm au}$ and $m=7~m_{\rm H}$. To rationalize this behavior  figure
\ref{fig:momentum} shows various expectation values for $t_f=14000~{\rm au}$ and $m=6$ and $7~m_{\rm H}$.
The first row compares Gaussian and quantum results and we can notice that the corresponding trajectories diverge considerably more for $7~m_{\rm H}$ (b) than for $6~m_{\rm H}$  (a), even though a naive consideration would suggest that the performance of the single Gaussian approximation is  better for the more massive  particle. But, we notice  a notable difference in variance of the Gaussian and quantum wavepackets as a likely reason for the discrepancy. In general we observe that while in the good performing cases the wavepacket essentially stay localized, the opposite is true for the poor performing cases. This holds irrespective of the actual mass of the particle. From the second and fourth rows of figure \ref{fig:momentum} we notice that  the cases $m=6$ and $7~m_{\rm H}$ differ in the momentum and thus kinetic energy when crossing the barrier. While in the former case the momentum is maximum at the barrier top, in the latter the particle is slowed down when reaching the barrier. As a consequence it becomes rather delocalized in position space and thus the single gaussian approximation fails.

\renewcommand{\figwidth}{0.9}
\begin{figure}[th]
\centering
        \includegraphics[width=\figwidth\linewidth]{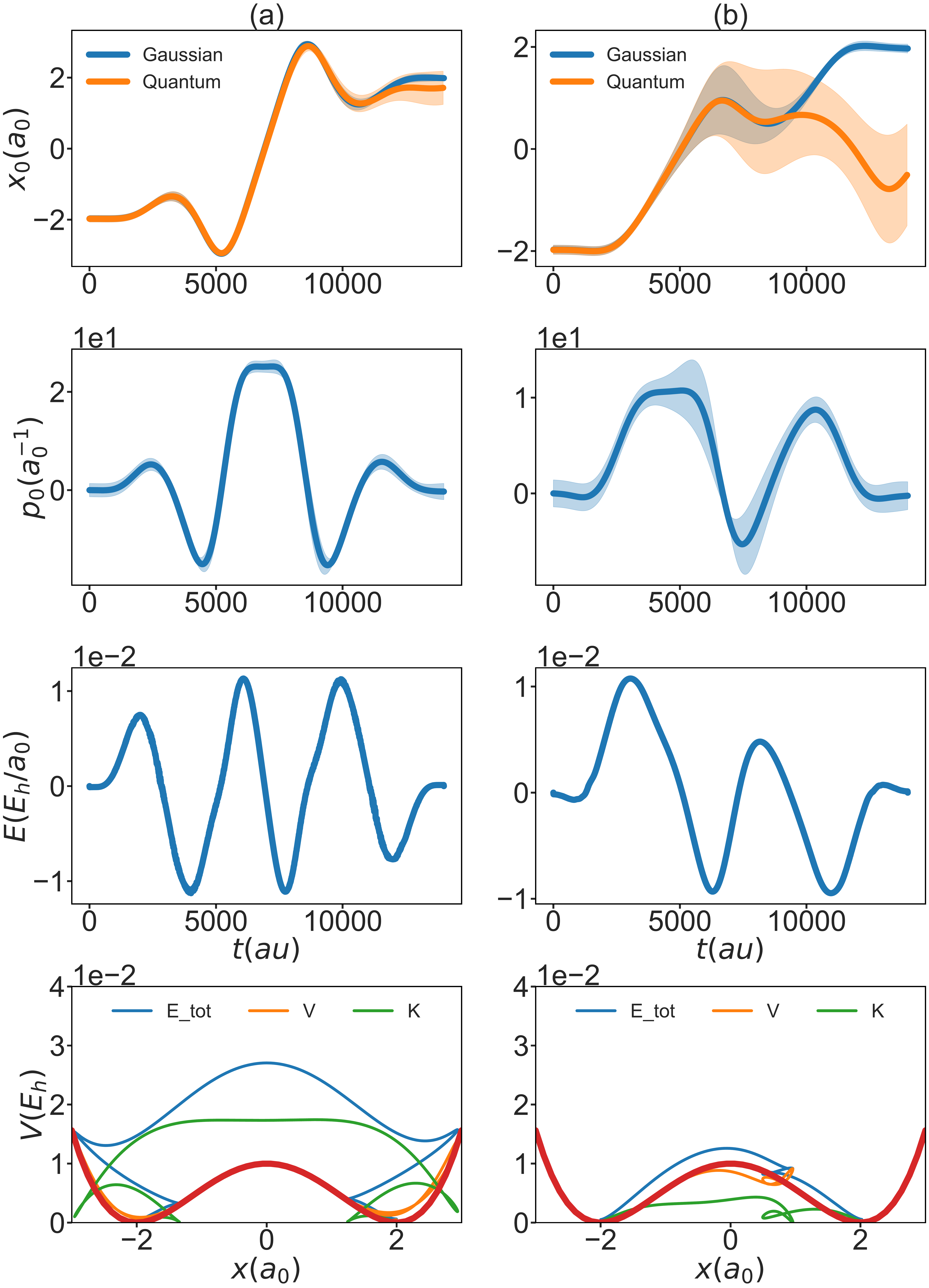}
        \caption{Expectation values of coordinate and  momentum, optimal field, as well as total, potential ($V$) and kinetic ($K$) energy of the moving wave packet (rows from top to bottom) for  $t_{\rm f}=14000~{\rm au}$ and  (a) $6~m_{\rm H}$, (b) $7~m_{\rm H}$. In the bottom row the expectation values  are plotted at the respective positions of the Gaussian wavepacket.}\label{fig:momentum}
\end{figure}
\clearpage

 In principle one could expect that decreasing the penalty factor $\kappa$ would alleviate this problem, i.e. stronger fields would imply higher momentum. However, after inspecting figure \ref{fig:momentum}, it is apparent that for a given final time it depends on the initial direction of momentum whether the wavepacket will pass the barrier with high or low momentum. This idea supports the conclusion that not only the mass of the particle, but also the specific optimal path, are important for the validity of the single Gaussian approximation. Controlling the initial direction in a way which works in a black-box fashion for all cases covered in figure \ref{fig:error} has not been successfull. However, in contrast to indirect control, where one would have to compute running cost derivatives with respect to state variables to get coupling terms between forward and backward Schr\"odinger equation, including additional running costs is straightforward in direct control. To demonstrate this we have added a second term to the running costs of equation \ref{run_cost}, which serves to maximize the kinetic energy, i.e.
\begin{equation}\label{run_cost_p0^2}
{\mathcal R}'[p_0(t),t] = - \eta \frac{p_0^2(t)}{2 m}~.
\end{equation}
Here, $\eta$ is a penalty scaling factor and the minus sign ensures that this term gets maximized. It is  expected that this will lead to barrier crossing with high momentum and thus a reduced error, equation (\ref{eq:error}). 

\renewcommand{\figwidth}{0.6}
\begin{figure}[th]
\centering
        \includegraphics[width=\figwidth\linewidth]{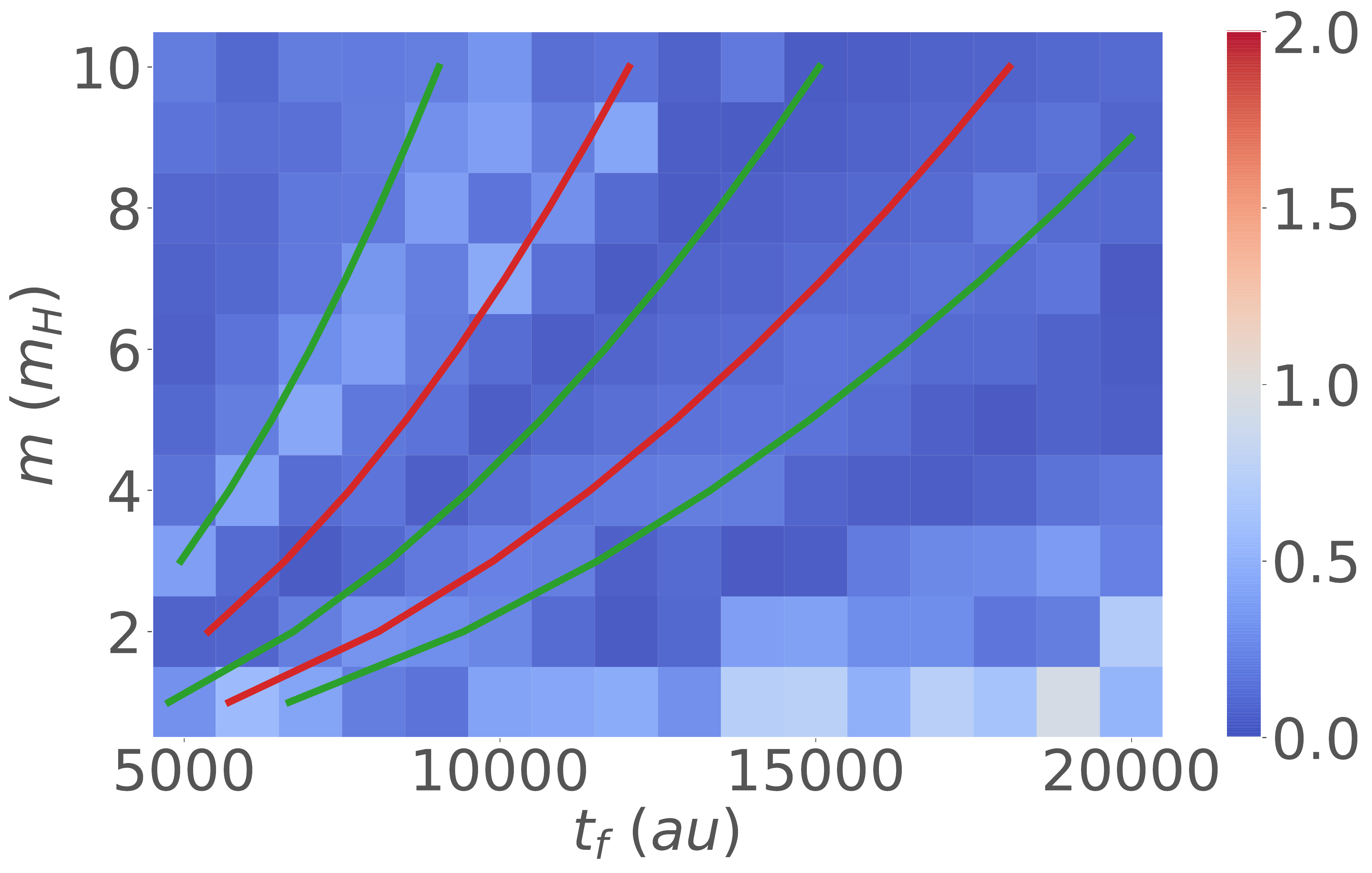}
        \caption{Error according to equation (\ref{eq:error}) as a function of different final times and masses. Running cost according to equation (\ref{run_cost_p0^2}) has been used together with equation (\ref{run_cost}). The penalty scaling factor was $\eta=0.003$, except for a few cases where lower or higher values has been used, ranging from $0.001$ to $0.015$.}\label{fig:error_p0^2}
\end{figure}
The results shown in figure \ref{fig:error_p0^2} clearly support our hypothesis, i.e. adding the running cost functional equation  (\ref{run_cost_p0^2}) leads to the elimination of the poor-performing islands. Hence, using the flexibility of the direct optimal control approach the region of validity of the single Gaussian approximation could be extended.
\section{Conclusions}
In this paper we have introduced a new tool for quantum optimal control. In contrast to indirect methods, which require the solution of a two-point boundary value problem, the present direct method builds on the first discretize and then optimize paradigm. Thus, by construction there is no need for explicit propagation of a wavepacket.
So far direct methods have found application mostly in engineering ~\cite{becerra10_1391,pardo16_946}. The performance and capabilities of the direct method have been demonstrated for the case of one-dimensional particle transfer in a bistable potential. For simplicity the wavepacket has been approximated by a single Gaussian function, but in principle other forms  are possible, e.g. superposition of Gaussians~\cite{richings15_269} or even expansions in terms of an eigenstate basis. Of course, Gaussians have the potential advantage of being suited for on-the-fly simulations, which brings OCT into the realm of the dynamics of complex molecular systems, at least in principle.

For a simple test system the question has been addressed whether the quantumness of the dynamics influences the final control yield, given a field which has been optimized for the single Gaussian approximation. Interestingly,  it turned out that nearly complete particle transfer can be achieved for a wide range of masses and final times. Here, the important point is whether the wavepacket crosses the barrier with high or low momentum, which for the given model is decided by the sign of the momentum during the initial dynamics. As a consequence, even the optimization based on a simple Gaussian wavepacket, possibly using on-the-fly dynamics, may provide reasonable control fields.

\section*{Conflict of Interest Statement}
The authors declare that the research was conducted in the absence of any commercial or financial relationships that could be construed as a potential conflict of interest.

\section*{Author Contributions}

ARR has performed the work and analyzed the results. OK has designed the project and supervised the scientific work. All authors have discussed and interpreted the results and contributed to writing the  manuscript.

\section*{Funding}
This work has been funded by the grant Ku952/10-1 from the  Deutsche Forschungsgemeinschaft (DFG).




\begin{thebibliography}{42}
\expandafter\ifx\csname natexlab\endcsname\relax\def\natexlab#1{#1}\fi
\expandafter\ifx\csname urlstyle\endcsname\relax
  \expandafter\ifx\csname doi\endcsname\relax
  \def\doi#1{doi:\discretionary{}{}{}#1}\fi \else
  \expandafter\ifx\csname doi\endcsname\relax
  \def\doi{doi:\discretionary{}{}{}\begingroup \urlstyle{rm}\Url}\fi \fi
\expandafter\ifx\csname selectlanguage\endcsname\relax
  \def\selectlanguage#1{}\fi

\bibitem[{Judson and Rabitz(1992)}]{judson92_1500}
Judson RS, Rabitz H.
\newblock Teaching lasers to control molecules.
\newblock {\em Phys. Rev. Lett.\/} {\bf 68} (1992) 1500.

\bibitem[{Paramonov and Savva(1983)}]{paramonov83_340}
Paramonov GK, Savva VA.
\newblock Resonance {{Effects}} in {{Molecule Vibrational}}-{{Excitation}} by
  {{Picosecond Laser}}-{{Pulses}}.
\newblock {\em Phys. Lett. A\/} {\bf 97} (1983) 340--342.

\bibitem[{Tannor and Rice(1985)}]{tannor85_5013}
Tannor DJ, Rice SA.
\newblock Control of {{Selectivity}} of {{Chemical Reaction}} via {{Control}}
  of {{Wave Packet Evolution}}.
\newblock {\em J Chem Phys\/} {\bf 83} (1985) 5013--5018.

\bibitem[{Tannor et~al.(1986)Tannor, Kosloff, and Rice}]{tannor86_5805}
Tannor DJ, Kosloff R, Rice SA.
\newblock Coherent pulse sequence induced control of selectivity of reactions:
  {{Exact}} quantum mechanical calculations.
\newblock {\em J Chem Phys\/} {\bf 85} (1986) 5805.

\bibitem[{Brumer and Shapiro(1986)}]{brumer86_541}
Brumer P, Shapiro M.
\newblock Control of unimolecular reactions using coherent light.
\newblock {\em Chem. Phys. Lett.\/} {\bf 126} (1986) 541--546.

\bibitem[{Shi et~al.(1988)Shi, Woody, and Rabitz}]{shi88_6870}
Shi S, Woody A, Rabitz H.
\newblock Optimal control of selective vibrational excitation in harmonic
  linear chain molecules.
\newblock {\em J. Chem. Phys.\/} {\bf 88} (1988) 6870--6883.

\bibitem[{Shi and Rabitz(1989)}]{shi89_185}
Shi S, Rabitz H.
\newblock Selective excitation in harmonic molecular systems by optimally
  designed fields.
\newblock {\em Chem. Phys.\/} {\bf 139} (1989) 185--199.

\bibitem[{Kosloff et~al.(1989)Kosloff, Rice, Gaspard, Tersigni, and
  Tannor}]{kosloff89_201}
Kosloff R, Rice S, Gaspard P, Tersigni S, Tannor D.
\newblock Wavepacket dancing: {{Achieving}} chemical selectivity by shaping
  light pulses.
\newblock {\em Chem. Phys.\/} {\bf 139} (1989) 201--220.

\bibitem[{Brif et~al.(2010)Brif, Chakrabarti, and Rabitz}]{brif10_075008}
Brif C, Chakrabarti R, Rabitz H.
\newblock Control of quantum phenomena: Past, present and future.
\newblock {\em New J. Phys.\/} {\bf 12} (2010) 075008.

\bibitem[{Werschnik and Gross(2007)}]{werschnik07_R175}
Werschnik J, Gross EKU.
\newblock Quantum optimal control theory.
\newblock {\em J. Phys. B: At. Mol. Opt. Phys.\/} {\bf 40} (2007) R175--R211.

\bibitem[{Worth and Richings(2013)}]{worth13_113}
Worth GA, Richings GW.
\newblock Optimal control by computer.
\newblock {\em Annu. Rep. Prog. Chem., Sect. C: Phys. Chem.\/} {\bf 109} (2013)
  113.

\bibitem[{Keefer and {de Vivie-Riedle}(2018)}]{keefer18_2279}
Keefer D, {de Vivie-Riedle} R.
\newblock Pathways to {{New Applications}} for {{Quantum Control}}.
\newblock {\em Acc. Chem. Res.\/} {\bf 51} (2018) 2279--2286.

\bibitem[{Brixner and Gerber(2003)}]{brixner03_418}
Brixner T, Gerber G.
\newblock Quantum {{Control}} of {{Gas}}-{{Phase}} and {{Liquid}}-{{Phase
  Femtochemistry}}.
\newblock {\em ChemPhysChem\/} {\bf 4} (2003) 418.

\bibitem[{Prokhorenko et~al.(2006)Prokhorenko, Nagy, Waschuk, Brown, Birge, and
  Miller}]{prokhorenko06_1257}
Prokhorenko VI, Nagy AM, Waschuk SA, Brown LS, Birge RR, Miller RJD.
\newblock Coherent {{Control}} of {{Retinal Isomerization}} in
  {{Bacteriorhodopsin}}.
\newblock {\em Science\/} {\bf 313} (2006) 1257--1261.

\bibitem[{Stensitzki et~al.(2018)Stensitzki, Yang, Kozich, Ahmed, K{\"o}ssl,
  K{\"u}hn et~al.}]{stensitzki18_126}
Stensitzki T, Yang Y, Kozich V, Ahmed AA, K{\"o}ssl F, K{\"u}hn O, et~al.
\newblock Acceleration of a ground-state reaction by selective
  femtosecond-infrared-laser-pulse excitation.
\newblock {\em Nat. Chem.\/} {\bf 10} (2018) 126--131.

\bibitem[{Nunes et~al.(2020)Nunes, Pereira, Reva, Amado, Cristiano, and
  Fausto}]{nunes20_8034}
Nunes CM, Pereira NAM, Reva I, Amado PSM, Cristiano MLS, Fausto R.
\newblock Bond-{{Breaking}}/{{Bond}}-{{Forming Reactions}} by {{Vibrational
  Excitation}}: {{Infrared}}-{{Induced Bidirectional Tautomerization}} of
  {{Matrix}}-{{Isolated Thiotropolone}}.
\newblock {\em J. Phys. Chem. Lett.\/}  (2020) 8034--8039.

\bibitem[{Heyne and K{\"u}hn(2019)}]{heyne19_11730}
Heyne K, K{\"u}hn O.
\newblock Infrared {{Laser Excitation Controlled Reaction Acceleration}} in the
  {{Electronic Ground State}}.
\newblock {\em J. Am. Chem. Soc.\/} {\bf 141} (2019) 11730--11738.

\bibitem[{Zhu and Rabitz(1998)}]{zhu98_385}
Zhu W, Rabitz H.
\newblock A rapid monotonically convergent iteration algorithm for quantum
  optimal control over the expectation value of a positive definite operator.
\newblock {\em J. Chem. Phys.\/} {\bf 109} (1998) 385--391.

\bibitem[{Kelly(2017)}]{kelly17_849}
Kelly M.
\newblock An introduction to trajectory optimization: {{How}} to do your own
  direct collocation.
\newblock {\em SIAM Rev.\/} {\bf 59} (2017) 849--904.

\bibitem[{Kappen(2007)}]{kappen07_149}
Kappen HJ.
\newblock An introduction to stochastic control theory, path integrals and
  reinforcement learning.
\newblock {\em AIP Conf. Proc.\/} {\bf 887} (2007) 149--181.

\bibitem[{{Chen-Charpentier} and Jackson(2020)}]{chen-charpentier20_112983}
{Chen-Charpentier} BM, Jackson M.
\newblock Direct and indirect optimal control applied to plant virus
  propagation with seasonality and delays.
\newblock {\em J. Comput. Appl. Math.\/} {\bf 380} (2020) 112983.

\bibitem[{Betts(2010)}]{betts10_}
Betts JT.
\newblock {\em Practical {{Methods}} for {{Optimal Control}} and {{Estimation
  Using Nonlinear Programming}}\/}.
\newblock Advances in {{Design}} and {{Control}} ({Society for Industrial and
  Applied Mathematics}) (2010).

\bibitem[{Pardo et~al.(2016)Pardo, Moller, Neunert, Winkler, and
  Buchli}]{pardo16_946}
Pardo D, Moller L, Neunert M, Winkler AW, Buchli J.
\newblock Evaluating {{Direct Transcription}} and {{Nonlinear Optimization
  Methods}} for {{Robot Motion Planning}}.
\newblock {\em IEEE Robot. Autom. Lett.\/} {\bf 1} (2016) 946--953.

\bibitem[{Meyer et~al.(1990)Meyer, Manthe, and Cederbaum}]{meyer90_73}
Meyer HD, Manthe U, Cederbaum LS.
\newblock The multi-configurational time-dependent {{Hartree}} approach.
\newblock {\em Chem. Phys. Lett.\/} {\bf 165} (1990) 73--78.

\bibitem[{Beck et~al.(2000)Beck, J{\"a}ckle, Worth, and Meyer}]{beck00_1}
Beck MH, J{\"a}ckle A, Worth GA, Meyer HD.
\newblock The multiconfiguration time-dependent {{Hartree}} method: {{A}}
  highly efficient algorithm for propagating wavepackets.
\newblock {\em Phys. Rep.\/} {\bf 324} (2000) 1--105.

\bibitem[{Schr{\"o}der et~al.(2008)Schr{\"o}der, {Carreon-Macedo}, and
  Brown}]{schroder08_850}
Schr{\"o}der M, {Carreon-Macedo} JL, Brown A.
\newblock Implementation of an iterative algorithm for optimal control of
  molecular dynamics into {{MCTDH}}.
\newblock {\em PhysChemChemPhys\/} {\bf 10} (2008) 850.

\bibitem[{Accardi et~al.(2009)Accardi, Borowski, and K{\"u}hn}]{accardi09_7491}
Accardi A, Borowski A, K{\"u}hn O.
\newblock Nonadiabatic {{Quantum Dynamics}} and {{Laser Control}} of {{Br$_2$}}
  in {{Solid Argon}}.
\newblock {\em J. Phys. Chem. A\/} {\bf 113} (2009) 7491--7498.

\bibitem[{Richings et~al.(2015)Richings, {I. Polyak}, {K.E. Spinlove}, {G.A.
  Worth}, {I. Burghardt}, and {B. Lasorne}}]{richings15_269}
Richings G, {I Polyak}, {KE Spinlove}, {GA Worth}, {I Burghardt}, {B Lasorne}.
\newblock Quantum dynamics simulations using {{Gaussian}} wavepackets: The
  {{vMCG}} method.
\newblock {\em Int. Rev. Phys. Chem.\/} {\bf 34} (2015) 269--308.

\bibitem[{Richings and Habershon(2018)}]{richings18_134116}
Richings GW, Habershon S.
\newblock {{MCTDH}} on-the-fly: {{Efficient}} grid-based quantum dynamics
  without pre-computed potential energy surfaces.
\newblock {\em J. Chem. Phys.\/} {\bf 148} (2018) 134116.

\bibitem[{Heller(2018)}]{heller18_}
Heller EJ.
\newblock {\em The {{Semiclassical Way}} to {{Dynamics}} and
  {{Spectroscopy}}\/} ({Princeton}: {Princeton University Press}) (2018).

\bibitem[{Kondorskiy and Nakamura(2005)}]{kondorskiy05_75}
Kondorskiy A, Nakamura H.
\newblock Semiclassical {{Formulation}} of {{Optimal Control Theory}}.
\newblock {\em J. Theor. Comput. Chem.\/} {\bf 04} (2005) 75--87.

\bibitem[{Kondorskiy et~al.(2005)Kondorskiy, Mil'nikov, and
  Nakamura}]{kondorskiy05_041401}
Kondorskiy A, Mil'nikov G, Nakamura H.
\newblock Semiclassical guided optimal control of molecular dynamics.
\newblock {\em Phys. Rev. A\/} {\bf 72} (2005) 041401.

\bibitem[{{Bona{\v c}i{\'c}-Kouteck{\'y}} and
  Mitri{\'c}(2005)}]{bonacic-koutecky05_11}
{Bona{\v c}i{\'c}-Kouteck{\'y}} V, Mitri{\'c} R.
\newblock Theoretical {{Exploration}} of {{Ultrafast Dynamics}} in {{Atomic
  Clusters}}: {{Analysis}} and {{Control}}.
\newblock {\em Chem. Rev.\/} {\bf 105} (2005) 11--66.

\bibitem[{Broeckhove et~al.(1988)Broeckhove, Lathouwers, Kesteloot, and
  Van~Leuven}]{broeckhove88_547}
Broeckhove J, Lathouwers L, Kesteloot E, Van~Leuven P.
\newblock On the equivalence of time-dependent variational principles.
\newblock {\em Chem. Phys. Lett.\/} {\bf 149} (1988) 547--550.

\bibitem[{Worth and Sanz(2010)}]{worth10_15570}
Worth GA, Sanz CS.
\newblock Guiding the time-evolution of a molecule: Optical control by
  computer.
\newblock {\em Phys. Chem. Chem. Phys.\/} {\bf 12} (2010) 15570.

\bibitem[{Sundermann and de~{Vivie-Riedle}(2000)}]{sundermann00_1896}
Sundermann K, de~{Vivie-Riedle} R.
\newblock Extensions to quantum optimal control algorithms and applications to
  special problems in state selective molecular dynamics.
\newblock {\em J Chem Phys\/} {\bf 110} (2000) 1896.

\bibitem[{Do{\v s}li{\'c}(2006)}]{doslic06_013402}
Do{\v s}li{\'c} N.
\newblock Generalization of the {{Rabi}} population inversion dynamics in the
  sub-one-cycle pulse limit.
\newblock {\em Phys. Rev. A\/} {\bf 74} (2006) 013402.

\bibitem[{Do{\v s}li{\'c} et~al.(1998)Do{\v s}li{\'c}, K{\"u}hn, and
  Manz}]{doslic98_292}
Do{\v s}li{\'c} N, K{\"u}hn O, Manz J.
\newblock Infrared laser pulse controlled ultrafast {{H}}-atom switching in
  two-dimensional asymmetric double well potentials.
\newblock {\em Ber. Bunsen Ges. Phys. Chem.\/} {\bf 102} (1998) 292--297.

\bibitem[{Doslic et~al.(2011)Doslic, {Abdel-Latif}, and
  K{\"u}hn}]{doslic11_411}
Doslic N, {Abdel-Latif} MK, K{\"u}hn O.
\newblock Laser {{Control}} of {{Single}} and {{Double Proton Transfer
  Reactions}}.
\newblock {\em Acta Chim Slov\/} {\bf 58} (2011) 411--424.

\bibitem[{Becerra(2010)}]{becerra10_1391}
Becerra VM.
\newblock Solving complex optimal control problems at no cost with {{PSOPT}}.
\newblock {\em 2010 {{IEEE International Symposium}} on {{Computer}}-{{Aided
  Control System Design}}\/} ({Yokohama, Japan}: IEEE) (2010), 1391--1396.

\bibitem[{W{\"a}chter and Biegler(2006)}]{wachter06_25}
W{\"a}chter A, Biegler LT.
\newblock On the implementation of an interior-point filter line-search
  algorithm for large-scale nonlinear programming.
\newblock {\em Math. Program.\/} {\bf 106} (2006) 25--57.

\bibitem[{Worth et~al.()Worth, Beck, J{\"a}ckle, and Meyer}]{mctdh85}
Worth GA, Beck MH, J{\"a}ckle A, Meyer HD.
\newblock The {{MCTDH Package}}, {{Version}} 8.2, (2000). {{H}}.-{{D}}.
  {{Meyer}}, {{Version}} 8.3 (2002), {{Version}} 8.4 (2007). {{O}}.
  {{Vendrell}} and {{H}}.-{{D}}. {{Meyer}}, {{Version}} 8.5 (2011), used
  {{Version}} 8.5.4, {{See}} {{http://mctdh.uni-hd.de/}}.

\end{thebibliography}
\end{document}